\documentstyle[11pt,a4wide,epsf]{article}

\newcommand{\pspicture}[1]{
\centerline{\setlength\epsfxsize{9.2cm}\epsfbox{#1}}}

\newcommand{\be}{\begin{equation}}
\newcommand{\ee}{\end{equation}}
\newcommand{\bea}{\begin{eqnarray}}
\newcommand{\eea}{\end{eqnarray}}

\begin{document}
\pagestyle{plain}
\begin{flushright}
DAMTP Preprint: 97-28\\
\end{flushright}
\vskip 20pt

\begin{center}
{\huge \bf $\Upsilon$-spectrum from NRQCD with Improved Action} \\[15pt]
{\sc UKQCD Collaboration} \\[10pt]
T. Manke \footnote{tm10010@damtp.cam.ac.uk},
I.T. Drummond\footnote{itd@damtp.cam.ac.uk},
R.R. Horgan\footnote{rrh@damtp.cam.ac.uk} ,
H.P. Shanahan\footnote{H.P.Shanahan@damtp.cam.ac.uk}\\[10pt]
{\sl DAMTP, University of Cambridge, Cambridge CB3 9EW, England}
\end{center}

\begin{abstract}
We explore the effect of higher order operators in the
non-relativistic formulation of QCD (NRQCD). 
We calculated masses in the $b \bar b$-spectrum using
quenched gauge configurations at $\beta = 6.0$ and two different
NRQCD actions which have been corrected to order $mv^4$ and $mv^6$.
The two-point functions are calculated in a gauge invariant fashion. We
find the general structure of the spectrum to be the same in the two cases. Using the
$1\bar P - 1{}^3S_1$ splitting we determine the inverse lattice spacings to be
2.44(4) GeV and 2.44(5) GeV for the $mv^4$ and $mv^6$ actions,
respectively. We do observe shifts in the spin splittings. The hyperfine
splitting is reduced by approximately 4 MeV, while the fine splitting
is down by up to 10 MeV, albeit with large statistical errors.
\end{abstract}

\section{Introduction}
Heavy quark systems have long been studied and much
experimental data has been accumulated which will eventually lead 
to tight constraints on the parameters of the Standard Model. 
To this end it is important to have accurate non-perturbative predictions
from QCD which can be compared with experiment. One such method is
Lattice QCD and  it is necessary to understand how heavy quark
systems can be treated in this framework.

A nonrelativistic approximation to QCD, NRQCD, was proposed \cite{heavy_bound}
to go beyond the static approximation in lattice calculations.
NRQCD has been remarkably successful in reproducing the spectrum of
heavy quark systems \cite{davies_first,ron_beta62,ron_radial,davies_prec} owing to
the fact that the quarks within such hadrons move
with velocities $v$ such that $v^2 \ll c^2$. Furthermore, NRQCD has the virtue of
retaining, at least approximatively, the quark dynamics and can
be considered an effective theory.
Computationally, such an approach is much faster to solve than the
inversion problem in fully relativistic QCD.
A systematic improvement program has been
developed to match lattice NRQCD to continuum QCD \cite{nrqcd_improved,morningstar}.
As an effective field theory the predictive power of NRQCD relies on
the control of higher dimensional operators.
 
In this paper we calculate the $b \bar b$ spectrum comparing 
two different NRQCD-actions with accuracies $O(mv^4)$ and  $O(mv^6)$.
In section 2 we give the details of our evolution equation and the
gauge invariant operators used. In section 3 we state the results and convert them
into dimensionful units. In the concluding section we compare our
results with other work in this area and experimental data.

\newpage
\section{NRQCD and Operators}
In NRQCD the inversion problem of the fermion matrix is an initial
value problem. We use the evolution equation 
along the Euclidean time-direction defined by: 

\bea
G({\bf x},t+1;y) &=& \left(1-\frac{aH_0}{2n}\right)^n U_t^{\dag}(x) \left(1-\frac{aH_0}{2n}\right)^n
\left(1-a\delta H \right) G({\bf x},t;y) ~~,~~~~~t \ge t_y + 1~~,\nonumber \\
G({\bf x},t_y+1;y)   &=& \left(1-\frac{aH_0}{2n}\right)^n U_t^{\dag}(x)
\left(1-\frac{aH_0}{2n}\right)^n S({\bf x;y})~~.
\label{eq:evol}
\eea
where
\bea
H_0 &=& -  \frac{\Delta^2}{2m_b} ~~, \nonumber \\
\delta H &=& - c_0 \frac{\Delta^4}{8m_b^3} - c_1 \frac{1}{2m_b} \sigma\cdot g{\bf B}
+ c_2 \frac{i}{8m_b^2}(\Delta \cdot g{\bf E} - g{\bf E} \cdot \Delta)
- c_3 \frac{1}{8m_b^2} \sigma \cdot ( \tilde{\Delta} \times g{\bf E} -
g{\bf E} \times \tilde{\Delta})\nonumber \\
& & - c_4 \frac{1}{8m_b^3}\{\Delta^2,\sigma \cdot g{\bf B}\}  
    - c_5 \frac{1}{64m_b^3}\{\Delta^2, \sigma \cdot ( \Delta \times
g{\bf E} - g{\bf E} \times \Delta)\} 
- c_6 \frac{i}{8m_b^3} \sigma \cdot g{\bf E} \times g{\bf E} \nonumber \\
& & - c_7 \frac{a\Delta^4}{16n m_b^2} + c_8 \frac{a^2 \Delta^{(4)}}{24m_b} ~~.
\label{eq:dh}
\eea

\noindent Here $S({\bf x,y})$ is the source at the first timeslice ($t=t_y$).
We have $S({\bf x,y}) = \delta^{(3)}({\bf x,y})$ for a single quark source at the origin, $y$, but we also 
propagate extended objects with the same evolution equation.
The operator $H_0$ is the leading kinetic term and $\delta H$ contains
the relativistic corrections and spin corrections. 
The last two terms
in  $\delta H$ are present to correct for lattice
spacing errors in temporal and spatial derivatives.
For the derivatives we give the following definitions which are
consistent with those of  \cite{davies_prec,nrqcd_improved}:
\bea
a\Delta_i G(x,y) &=& U_i(x)G(x+\hat{\i},y) - U_{-i}(x)G(x-\hat{\i},y)=
\frac{1}{2}\left(e^{a\partial_i} - e^{-a\partial_i}\right)G(x,y)~~, \nonumber \\
\tilde{\Delta}_i &=& \Delta_i -
\frac{a^2}{6}\Delta_i^{+}\Delta_i^{\pm}\Delta_i^{-} = \partial_i +
O(a^4) ~~,\nonumber \\
a^2\Delta^2 G(x,y) &=& \left(\sum_{i=1}^{3} U_i(x)G(x+\hat{\i},y) +
U_{-i}(x)G(x-\hat{\i},y)\right) - 6~G(x,y) ~~,\nonumber \\
\Delta^4 G(x,y) &=& \Delta^2( \Delta^2 G(x,y)) ~~,\nonumber \\
a^2\Delta^{(4)} G(x,y) &=& \left(\sum_{i=\pm 1}^{\pm 3}
U_i(x)U_i(x+\hat{\i})G(x+2\hat{\i},y) - 4 U_i(x)G(x+\hat{\i},y)  + \frac{c}{2}~G(x,y) \right)~~,
\nonumber \\ c &=& (4+\frac{2}{u_0^2})~~.
\eea
For our calculation with accuracy $O(mv^4)$ we set $c_4,c_5$ and $c_6$
to zero and replace $\tilde \Delta$ by $\Delta$.
In this case we calculate the fields $\bf E$ and $\bf B$ from the clover field
$F_{\mu \nu}$ in the standard fashion \cite{nrqcd_improved}.
To determine the spin splittings with accuracy $O(mv^6)$ we retain the above coefficients
and the improved operator $\tilde \Delta$. Correspondingly, we replace standard
discretized gauge field term $F_{\mu \nu}$ by
\be
\tilde{F}_{\mu\nu} = \frac{5}{3}F_{\mu\nu} -
\frac{1}{6}\left(U_{\mu}(x)F_{\mu\nu}(x+\mu) U_{\mu}^{\dag}(x)+
U_{-\mu}(x)F_{\mu\nu}(x-\mu) U_{-\mu}^{\dag}(x) - (\mu
\leftrightarrow \nu)
\right)~~,
\ee
when this accuracy is needed for ${\bf E}$ and ${\bf B}$ in Equation \ref{eq:dh}.

All the gauge links are tadpole improved with $u_0=\langle 0|\frac{1}{3}Tr
U_{\mu\nu}|0 \rangle^{1/4}$ as suggested in \cite{viability_pt}. 
It has been demonstrated that tadpole improvement is crucial to
reproduce the spin splittings accurately, which are otherwise
underestimated. We assume that radiative corrections are sufficiently
accounted for by tadpole improvement and we therefore set all the
renormalisation coefficients, $c_i$, to 1.

To extract masses we calculate two-point functions of operators
with the appropriate quantum numbers.
In a non-relativistic setting gauge-invariant meson operators can be
constructed from the two-spinors $\chi^{\dag}(x)$ and $\Psi(y)$ (which represent
the antiquark field and quark field) and a Wilson line, $W(x,y)$.
Such operators have the following behaviour under rotation, parity
and charge conjugation
\bea
\chi^{\dag}(x) W(x,y) \Psi(y) &\stackrel{\cal R}{\longrightarrow}& \chi^{\dag}(Rx) W^{\cal R}(Rx,Ry) \Psi(Ry)~~,\nonumber\\
\chi^{\dag}(x) W(x,y) \Psi(y) &\stackrel{\cal P}{\longrightarrow}& - \chi^{\dag}(x_p) W^{\cal P}(x_p,y_p) \Psi(y_p)~~, \nonumber\\
\chi^{\dag}(x) W(x,y) \Psi(y) &\stackrel{\cal C}{\longrightarrow}& -
\Psi^T(x) W^{\cal C}(x,y) \chi^{\ast}(y) ~~.
\label{eq:transform}
\eea

\noindent The construction of spatially extended operators on the lattice has been reported
elsewhere \cite{cm_oper}. Here we adopt the following notation:
\bea
L_i^n(x) &\equiv& U_i(x)U_i(x+i) \ldots U_i(x+(n-1)i) = L_{-i}^{n \dag}(x+in)~~, \nonumber \\
\Delta_i^n \Psi(x) &\equiv& L_i^n(x)\Psi(x+ni) - L_{-i}^n(x)\Psi(x-ni)~~.
\label{eq:derivative}
\eea
With the generalised link variables of equation \ref{eq:derivative}
and the transformation properties of equation \ref{eq:transform}
we can construct operators with definite
$J^{PC}$, where $J$ labels the irreducible representations of
the octahedral group ($J=A_1,A_2,T_1,T_2,E$).

In order to increase the overlap between the meson and the ground states we
create on the lattice we use extended versions of the derivative as
defined in equation \ref{eq:derivative}.
The operators we used are listed in Table \ref{tab:operators}.
\noindent The meson correlator is written as a Monte Carlo average over all configurations
\be
{\rm C}^{nm}(x,y) = \langle {\bf tr}\left[G^{\dag}(x,y)G^{nm}(x,y)\right] \rangle~~,
\label{eq:correlator}
\ee
where ${\bf tr}$ denotes contraction over all internal degrees
of freedom and $G^{nm}$ is the smeared propagator defined by 
\be
G^{nm}(x,y) \equiv \sum_{z_1,z_2} O^n(x,z_1)G(z_1,z_2)O^{m \dag}(z_2,y)~~.
\ee
Here $(n,m)$ stands for the radii at (sink, source) and
$O^n$ is an operator as defined in Table \ref{tab:operators} but
with extended symmetric derivative $\Delta^n$.
For the smeared propagator we solve equation \ref{eq:evol} with
$S({\bf x,y}) = O^{m\dag}({\bf x,y})$
and multiply with $O^n$ at the sink.
We fix the origin at some (arbitrary) lattice
point, $y$, and sum over all spatial ${\bf x}$ so
as to project out the zero momentum mode. In addition we sum over
all polarisations to increase the statistics. 

\section{Simulation and Results}
We use quenched gauge field configurations at $\beta=6.0$ on a $16^3
\times 48$ lattice generated with the standard Wilson action 
at the EPCC in Edinburgh.
The propagators were calculated at the HPCF in Cambridge and at the
Hitachi Europe Ltd. High Performance Computing Center in
Maidenhead.
As the operators chosen are gauge invariant, gauge fixing is
not necessary. We used the extended operators defined in section 2 with
radii 1,2 and 3 in both the source and sink thus giving rise to a
$3 \times 3$ matrix of correlators for each set of quantum numbers.
The calculation were done with a bare quark mass of $am_b=1.71$ which
is the same as that used in \cite{davies_prec}.
We used $n=2$ as the stability parameter in Equation \ref{eq:evol}. 
The tadpole improvement coefficient appropriate to $\beta=6.0$ is
$u_0 = 0.878$ as in \cite{davies_prec}. We calculated quark
propagators for both actions from 8 points  $(i,j,k)$ where $i,j,k$ is
$0$ or $7$ on the first timeslice for each of the 499
configurations. It has been noted previously \cite{davies_prec} that such an arrangement
gives independent and uncorrelated measurements due to the small size
of the $b \bar b$-system on our lattice. Thus we have a total of 35928
correlators for each meson we considered (${}^1S_0, {}^3S_1,{}^1P_1,
{}^3P_{0,1,2}$). We also average over all polarisations and all spin components.

\begin{figure}
\pspicture{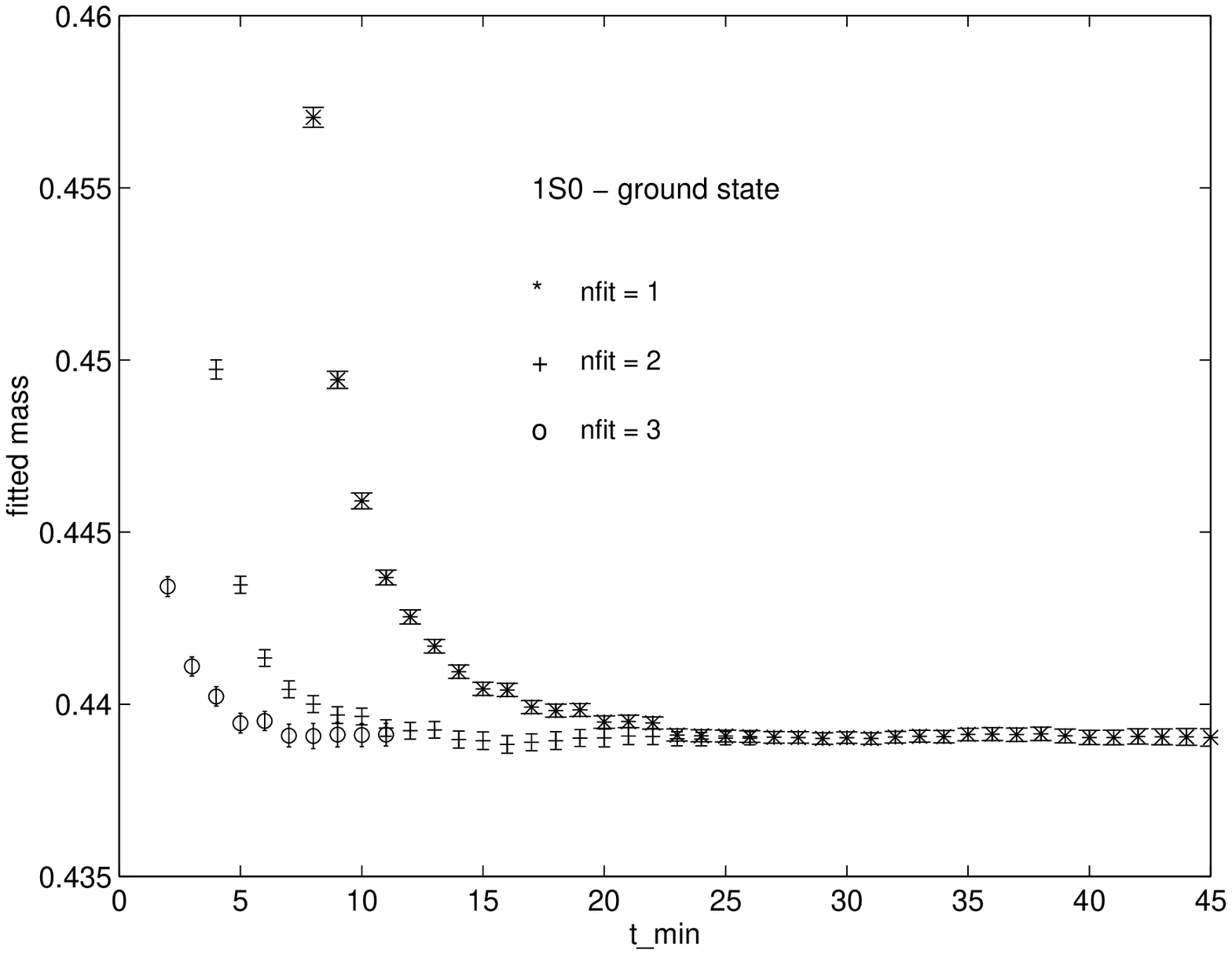}
\pspicture{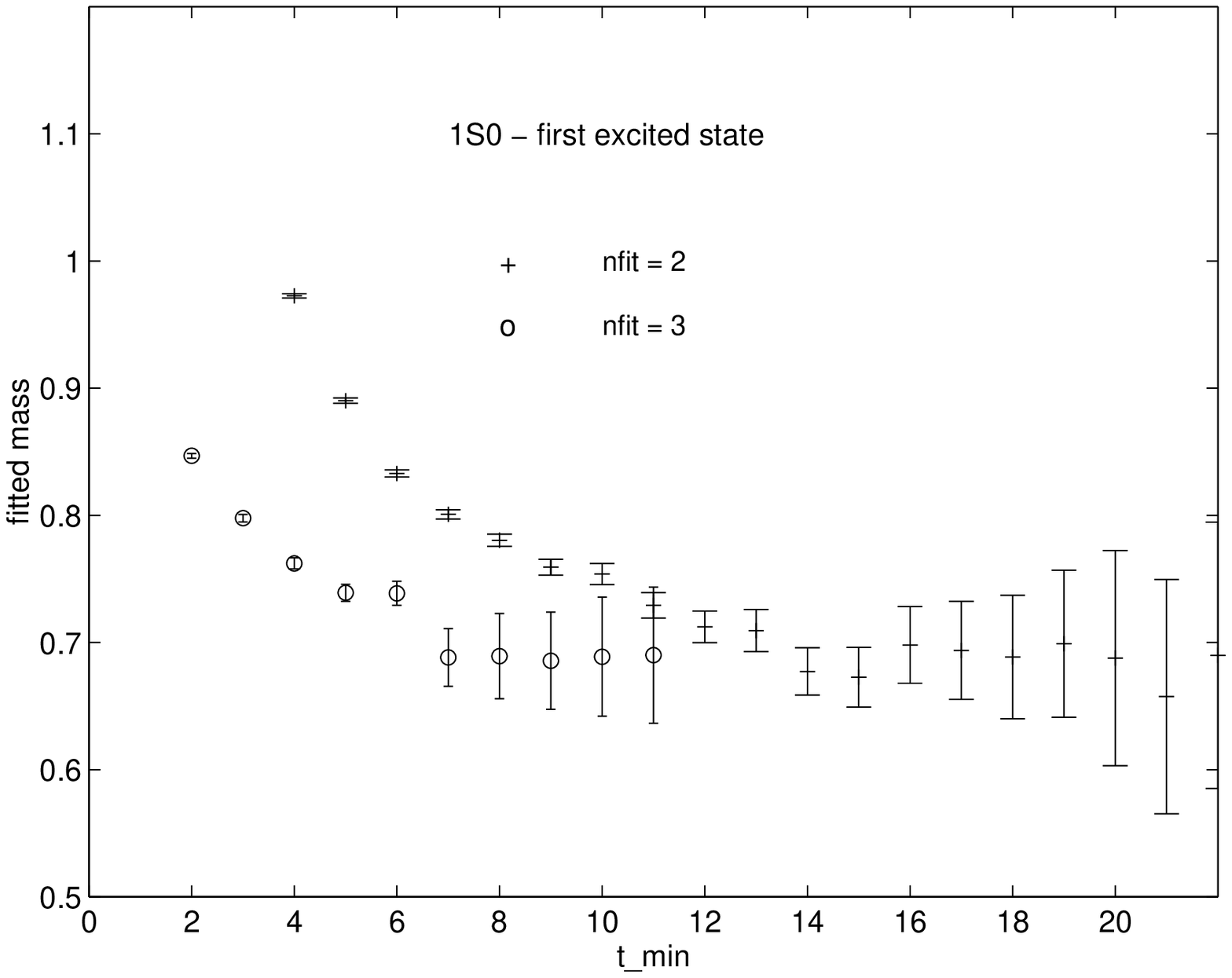}
\caption{Two examples of correlated fits for the
    $^1S_0$-state, calculated with the improved action.
    The different symbols denote different choices for $n_{fit}$ and
    the $t_{min}$ dependence is shown. This is to demonstrate the
    consistency of our fit results as we change the fit prescription.
    We chose $t_{max}=48$ for all fits.}
  \label{fig:correlated_fit}
\end{figure}

We fit the correlators to the multi-exponential form
\be
C_\alpha^{nm}(t) = \sum_{i=1}^{\rm n_{fit}} a_{\alpha i}^{nm} e^{-M_i^{\alpha} t}~~.
\label{eq:theory}
\ee
Here $\alpha$ denotes a meson state with certain quantum numbers, $(n,m)$ the
different radii at the (sink, source) and
$t$ the Euclidean time. We chose ${\rm n_{fit}}$ to be 1,2 or 3 and 
restrict ourselves to some conveniently chosen subsets of all the data,
e.g. a window $[t_{min},t_{max}]$. 
The basic method for obtaining the parameters $a_{\alpha i}^{nm}$ and
$M_i^{\alpha}$ (which we generically refer to as $\{b_l\}$) is a correlated
fit using the whole covariance matrix.
The statistical nature of the data causes fluctuations from the
central values of the parameters, which are determined from a minimal $\chi^2(b)$ fit.
We estimate the covariance matrix for the parameters from the inverse
of  $(\partial ^2 \chi^2)/(\partial b_k \partial b_l)$.
The goodness of the fit is quantified by the $Q$-value, defined
in \cite{numrec}. We require an acceptable fit to have $Q > 0.1$
and that the results remain consistent as we alter our
fitting prescription slightly (e.g. varying $t_{min}$ or the number
of eigenvalues retained when inverting the covariance matrix of the
data).
To illustrate the method we display a fitted mass
plotted against $t_{min}$ in Figure \ref{fig:correlated_fit}.
Our dimensionless results for the non-relativistic rest energies 
are shown in Table \ref{tab:compare_absolut}.

Fitting excitation energies relative to the ground state makes it
difficult to extract the splitting and the associated error of
the P-state. We rewrite Equation \ref{eq:theory} as
\be
C_{\alpha}^{nm}(t) = \sum_{i=1}^{n_{\rm fit}} a_{\alpha i}^{nm} e^{-(M_1^1+\delta{M}_i^\alpha) t}~~,
\label{eq:theory_split}
\ee
where we can now fit the amplitudes $a_{\alpha i}^{nm}$, the ground state mass $M_1^1$ and
the splittings, $\delta{M}_i^\alpha$, with respect to $M_1^1$. 
Our results are tabulated in Table 
\ref{tab:compare_split}.

\begin{figure}[t]

  \begin{center}
    \begin{tabular}{c}
    \hbox{\epsfxsize = 10cm  \epsffile{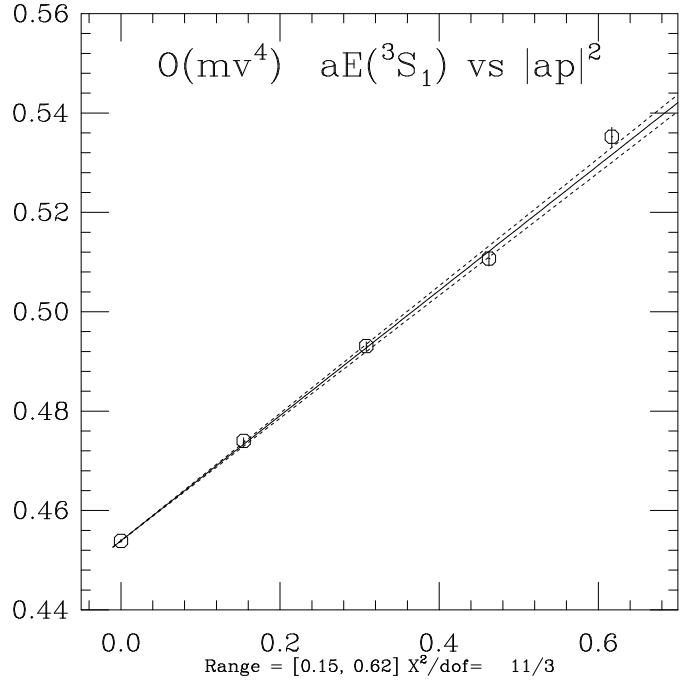}}\\
    \hbox{\epsfxsize = 10cm  \epsffile{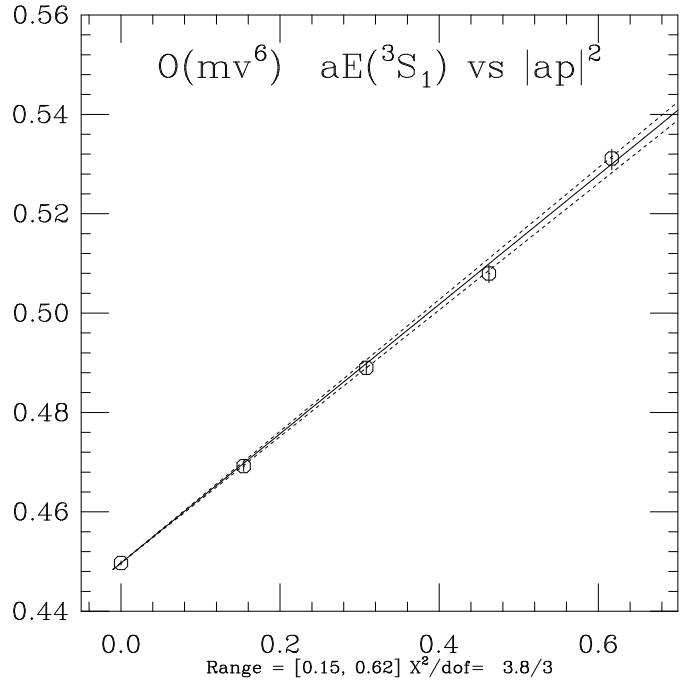}}
    \end{tabular}
  \end{center}

  \caption{Dispersion relation for $^3S_1$ and two different accuracies. The
  non-relativistic energy in lattice units is plotted
  vs. $(\frac{2\pi}{16})^2 |{\bf n}|^2$ where ${\bf
  n}=(1,0,0), {\bf n}=(1,1,0), {\bf n}=(1,1,1)$. From this we
  determine its kinetic mass to be $9.7(3)$ GeV ($O(mv^4)$) and
  $9.4(3)$ GeV ($O(mv^6)$) for $am_b=1.71$.}
  \label{fig:dispersion}
\end{figure}

From the $1\bar P-1{}^3S_1$ splitting we estimate the lattice spacing at
$\beta=6.0$ to be $a_{\bar P - S}^{-1}=2.44(4)$ GeV for the improved
action and $a_{\bar P - S}^{-1}=2.44(5)$ GeV for the unimproved
action. The kinetic masses were determined using 80
configurations for both actions.
We show an example of the dispersion relation in
Figure \ref{fig:dispersion} and the results are shown in Table
\ref{tab:dispersion}. 

\newpage

\section{Conclusions}
\begin{figure}[t]

  \begin{center}
    \begin{tabular}{c}
    \hbox{\epsfxsize = 10cm  \epsffile{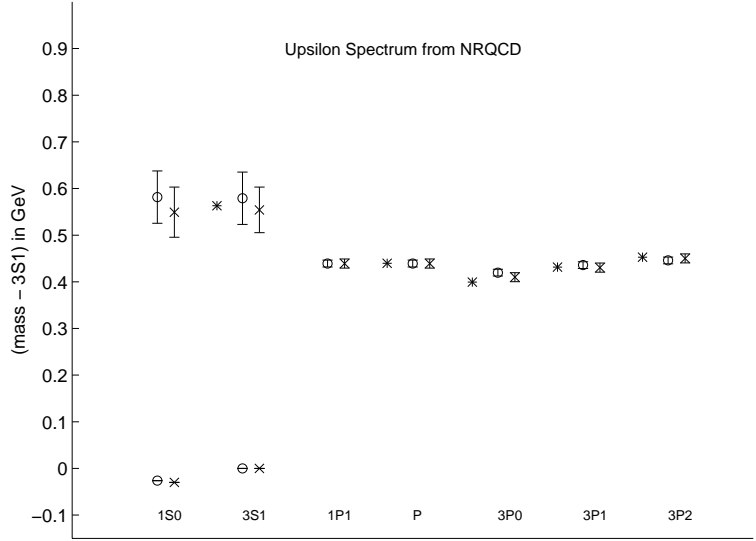}}
    \end{tabular}
  \end{center}

  \caption{Upsilon spectrum. We compare our results for two actions
  with different accuracies. Crosses are the results from an action
  which is accurate to $O(mv^4)$. Circles denote an
  improved calculation where spin-corrections, $c_4,c_5$ and $c_6$,
  have been taken into account up to order $O(mv^6)$. Bursts denote
  the experimental value where available. The
  splittings, relative to the $^3S_1$, are shown. The scale has been
  calculated from the $1\bar{P} - 1 {}^3S_1$ splitting.}
  \label{fig:spectrum}
\end{figure}

\begin{figure}[ht]

  \begin{center}
    \begin{tabular}{c}
    \hbox{\epsfxsize = 10cm  \epsffile{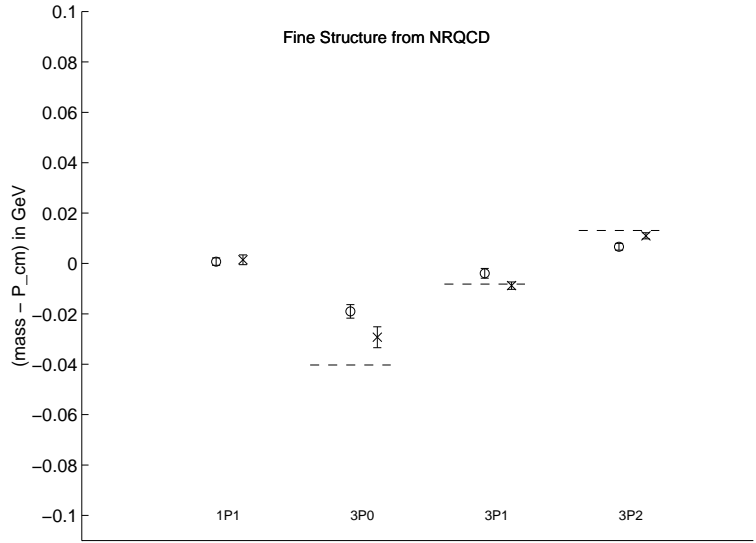}}
    \end{tabular}
  \end{center}

  \caption{Upsilon fine structure.
  The splittings, relative to the $\bar P$, are shown. 
  Experimental values are shown as dashed lines. Crosses denote
  accuracy up to $O(mv^4)$, circles up to $O(mv^6)$.
  The scale has been calculated from the $1\bar{P} - {}^3S_1$ splitting.}
  \label{fig:finestructure}
\end{figure}

Despite the success of NRQCD, a source of concern has been the 
effect of higher dimensional operators in ${\cal L}_{NRQCD}$. 
To test this issue, we calculated the $\Upsilon$ spectrum
in a gauge invariant fashion to accuracy $O(mv^4)$. We then retained
the spin corrections in Equation \ref{eq:evol} up to $O(mv^6)$ and
compared the results with each other.
We also checked higher order contributions to the derivative in the
term proportional to $c_3$ and found their effect negligible. Our main
results are summarised in Figures \ref{fig:spectrum} and
\ref{fig:finestructure}. In Table \ref{tab:compare_split} 
we compare the results for the two different
evolution equations with each other and convert the dimensionless
numbers into physical units. We set the scale using the
$1 \bar P - 1 {}^3S_1$ splitting and find $a^{-1}_{\bar P -
 S}=2.44(4)$ GeV and $2.44(5)$ GeV, respectively. 
Our results for a lower order evolution equation match those
from an earlier calculation \cite{davies_prec,davies_japan,davies_preliminary} 
where the configurations were fixed in the Coulomb gauge.

These comparisons demonstrate the consistency of NRQCD.
The gross structure of the $\Upsilon$-spectrum calculated is
unchanged when we added the spin correction terms. For example, the
ratio $R_{SP}=(2 {}^3S_1 - 1 {}^3S_1)/(1 \bar P - 1 {}^3S_1)=1.3(1)$
agrees for both actions and with the experimental value, 1.28, within
statistical errors.  

The bare mass parameter, $am_b$, was chosen to be 1.71 to
match a choice in \cite{davies_prec}. With our lattice spacing this
corresponds to $m_b=4.17(7)$ GeV and gives a kinetic mass of $9.4(3)$
GeV (for the $mv^6$ action) and $9.7(3)$ GeV (for the $mv^4$ action)  which
should be compared to the experimental value of $9.46$ GeV.
The differences in the kinetic mass due to the extra terms is
of the order of 1-2 standard deviations.
We note also that the $\chi^2/$dof resulting from the kinetic mass
fits for the $mv^6$ action is much smaller than that for the $mv^4$
action.

However, differences can be seen for the spin splittings.
The introduction of the extra terms reduces the hyperfine splitting,
${}^3S_1 - {}^1S_0$, by approximately 4 MeV. This size of a shift
is what one would expect from power counting arguments. 

We also observe a reduction in the fine structure. For example, by adding the
spin corrections the ${}^3P_2 - {}^3P_1$ splitting is reduced from
19(3) MeV to 11(3) MeV, compared with the experimental value of 21
GeV. Such a behaviour has also been reported for the spectrum calculations
of charmonium on much coarser lattices \cite{trottier}. Here we find a
similar tendency and further analysis will be needed to decide whether
this feature persists in unquenched calculations. We note that the ratio $R_{fs}=(1
{}^3P_2 - 1 {}^3P_1)/(1 {}^3P_1 - 1 {}^3P_0) = 0.56(19)$ from our
improved calculation is in better agreement with the experimental
value (0.66). For the less accurate action we find 1.11(26) for this ratio.

There will be additional uncertainties due to radiative corrections to
the coefficients $c_i$ in ${\cal L}_{NRQCD}$.
Discretisation errors in the gluonic action will also effect our
results and of course there will be corrections due to the
introduction of sea quarks.
Future studies into the accuracy of NRQCD will have to address these
systematic errors.

\newpage
\begin{center}{\bf Acknowledgements}\end{center}
We would like to thank other members of the UKQCD collaboration in
particular C.T.H. Davies and S. Collins for useful discussions.
T.M. wishes to thank J. Sloan for discussing details of the evolution
equation. This research is supported by a grant from EPSRC (Ref. No. 94007885) 
and in parts by a grant from the European Union (ERBCHB6-CT94-0523).
Our calculations are performed on  the Hitachi (S3600 and SR2001) located at 
Hitachi Europe's Maidenhead (United Kingdom) headquarters, the
University of Cambridge High Performance Computing Facility and the
CRAY-T3D at the Edinburgh Parallel Computing Centre at Edinburgh University.

\newpage

\begin{table}[h]
\begin{center}
\begin{tabular}{|c|c|c|c|c|}
\hline
Meson & $L\to O_h$ & $S \to O_h$ & $(L \otimes S)$ & Operator  \\
\hline
$^1S^{-+}$ & $0\to A_1$ & $0 \to A_1$ & $A_1^{-+}$&$c+\Delta^2$\\
$^3S^{--}$ & $0\to A_1$ & $1 \to T_1$ & $T_1^{--}$&$(c+\Delta^2)\sigma_i$ \\
\hline
$^1P^{+-}$ & $1 \to T_1$ & $0 \to A_1$ & $T_1^{+-}$ & $\Delta_i$\\
$^3P^{++}$ & $1 \to T_1$ & $1 \to T_1$ & $A_1^{++}$ & $\sigma_i \cdot \Delta_i$ \\
& & & $T_1^{++}$ & $\epsilon_{ijk}\sigma_j\Delta_k$ \\
& & & $T_2^{++}$ & $\sigma_k\Delta_j + \sigma_j\Delta_k$ \\
& & & $E^{++}$   & $\sigma_j\Delta_j - \sigma_k\Delta_k$ \\
\hline
\end{tabular}
\end{center}
\caption{transformation behaviour of $b\bar b$-states in their
non-relativistic notation}
\label{tab:operators}
\end{table}

\begin{table}[ht]
\begin{center}
\begin{tabular}{|l|l|l|l|}
\hline
State & $aM_{mv^6}$ & $aM_{mv^4}$ & $aM_{mv^4}-aM_{mv^6}$\\
\hline
\hline
$1 ^{1}S_0$ & 0.4391(3) & 0.4416(3) & 0.00265(5) \\
$2 ^{1}S_0$ & 0.688(23) & 0.679(22) & -- \\
\hline
$1 ^{3}S_1$ & 0.4497(4) & 0.4539(3) & 0.00439(6)\\
$2 ^{3}S_1$ & 0.687(23) & 0.681(20) & -- \\
\hline
$1 S$ & 0.4470(3) &  0.4508(3) & 0.00396(9) \\
$2 S$ & 0.69(3)   &  0.68(2)   & -- \\
\hline
$^1P_1$  & 0.632(7) & 0.635(7) & 0.0033(1)\\
$^3P_0$  & 0.619(7) & 0.621(7) & 0.0021(2)\\
$^3P_1$  & 0.628(7) & 0.627(8) & 0.0029(1)\\
$^3PT_2$ & 0.635(5) & 0.642(6) & 0.0053(1)\\
$^3PE_2$ & 0.632(7) & 0.642(6) & 0.0055(1)\\
\hline
$\bar P$ & 0.630(3)  & 0.634(4)  & 0.0042(1) \\
\hline
\end{tabular}
\end{center}
\caption{Here we compare the absolute masses from different
evolution equations and calculate their difference from a ratio fit
with bootstrap errors.}
\label{tab:compare_absolut}
\end{table}

\begin{table}
\begin{center}
\begin{tabular}{|l|l|l|l|l|l|}
\hline
Splitting & $O(mv^6)$ & $\times ~~a^{-1}$ [GeV]& $O(mv^4)$  & $\times
~~a^{-1}$ [GeV] & Exp. value\\
      &           & [2.44(4)] &   & [2.44(5)] & \\
\hline
$\bar{P} - 1 {}^3S_1$   & 0.180(3)   & 0.4398  & 0.180(4) & 0.4398  & 0.4398 GeV\\
\hline 
$1 ^{3}S_1 - 1 {}^1S_0$ & 0.01073(5) & 0.0262(6) & 0.01227(11)   & 0.0300(7)   & -- \\
$2 ^{3}S_1 - 1 {}^3S_1$ & 0.24(2)    & 0.59(5)   & 0.23(2)     & 0.56(5)   & 0.5629 GeV\\
$1 ^{1}P_1 - 1 {}^3S_1$ & 0.180(6)   & 0.440(18) & 0.181(7)    & 0.442(20) & -- \\
$1 ^{3}P_2 - 1 {}^3P_0$ & 0.011(2)   & 0.027(5)  & 0.0143(34)  & 0.035(8)  & 0.0534 GeV\\
$1 ^{3}P_2 - 1 {}^3P_1$ & 0.0045(14) & 0.011(3)  & 0.0078(11)  & 0.019(3)  & 0.0213 GeV\\
$1 ^{3}P_1 - 1 {}^3P_0$ & 0.008(1)   & 0.019(2)  & 0.0070(13)  & 0.017(3)  & 0.0321 GeV\\
\hline
$1 ^{3}P_2 - \bar{P}$ & 0.00272(55) & 0.0066(14)  & 0.0045(5)  & 0.011(1)   & 0.0131 GeV \\
$\bar{P} - 1 ^{3}P_1$ & 0.0016(8)   & 0.004(2)    & 0.0036(6)  & 0.0088(15) & 0.0082  GeV\\
$\bar{P} - 1 ^{3}P_0$ & 0.0078(11)  & 0.019(3)    & 0.0120(17) & 0.029(4)   & 0.0403 GeV \\
\hline
\hline
$M_{kin}$ & 3.84(8) & 9.38(25)  & 3.97(8)  & 9.70(29) & 9.4604 GeV \\
\hline
$R_{SP}$ & 1.33(11) &   & 1.28(11)  &  & 1.2802 \\
\hline
$R_{fs}$ & 0.56(19) &   & 1.11(26)  &   & 0.6636 \\
\hline
\end{tabular}
\end{center}
\caption{Summary of splittings. We show our results for both
accuracies and convert them into physical units. 
The lattice spacing is determined from the $1 \bar P - 1 {}^3S$ splitting in both
cases. $R_{SP} = (2 ^3S_1 - 1 {}^3S_1)/(1 \bar P - 1 {}^3S_1)$,
$R_{fs} = ({}^3P_2 - {}^3P_1)/({}^3P_1 - {}^3P_0)$.} 
\label{tab:compare_split}
\end{table}

\begin{table}
\begin{center}
\begin{tabular}{|l|r|r|l|}
\hline
State & $aE_{mv^6}$ & $aE_{mv^4}$ & $aE_{mv^4}-aE_{mv^6}$\\
\hline
\hline
$1 ^{1}S_0({\bf p}=(1,0,0))$ & 0.458(1) & 0.462(1) & 0.00236(7) \\
$1 ^{1}S_0({\bf p}=(1,1,0))$ & 0.478(2) & 0.481(2) & 0.00212(8)\\
$1 ^{1}S_0({\bf p}=(1,1,1))$ & 0.496(4) & 0.499(3) & 0.00191(10)\\
$1 ^{1}S_0({\bf p}=(2,0,0))$ & 0.522(4) & 0.524(5) & 0.0018(3)\\
\hline
$aM_{kin}$                    & 3.77(7)   & 3.93(7) &  \\
\hline
\hline
$1 ^{3}S_1({\bf p}=(1,0,0))$ & 0.469(1) & 0.474(2) & 0.00427(5) \\
$1 ^{3}S_1({\bf p}=(1,1,0))$ & 0.489(2) & 0.493(2) & 0.00416(6) \\
$1 ^{3}S_1({\bf p}=(1,1,1))$ & 0.508(2) & 0.511(3) & 0.00406(7) \\
$1 ^{3}S_1({\bf p}=(2,0,0))$ & 0.531(5) & 0.535(5) & 0.00417(8) \\
\hline
$aM_{kin}$                    & 3.84(8)   & 3.97(8) &  \\
\hline
\end{tabular}
\end{center}
\caption{The dispersion relation for $^1S_0$ and $^3S_1$.
The momentum is given in lattice units of \mbox{${\bf p}=\frac{2\pi}{L}(i,j,k)$}
where $L=16$. The last column is the difference for both evolution
equations as obtained from a ratio fit.}
\label{tab:dispersion}
\end{table}

\newpage


\newpage

\end{document}